\documentclass[pre,superscriptaddress,nofootinbib,floatfix,twocolumn]{revtex4}
\usepackage[english]{babel}
\usepackage{color}
\usepackage{amsmath}
\usepackage{amsfonts}
\usepackage{amsbsy}
\usepackage{mathrsfs}
\usepackage{graphicx}  
\usepackage{subfigure}
\newcommand{\eeq}{\end{equation}}
\newcommand{\bea}{\begin{eqnarray}}
\newcommand{\eea}{\end{eqnarray}}

\def\lsim{\mathrel{\rlap{
\lower4pt\hbox{\hskip-3pt$\sim$}}
    \raise1pt\hbox{$<$}}}     %less than approx. symbol
\def\gsim{\mathrel{\rlap{
\lower4pt\hbox{\hskip-3pt$\sim$}}
    \raise1pt\hbox{$>$}}}     %greater than or approx. symbol

\begin{document}
\title{Exchange--correlation bound states of the triplet soft--sphere fermions by the
	path integral Monte Carlo simulations}
\author{V.S.~Filinov}
\author{R.A.~Syrovatka}
\author{P.R.~Levashov}
\affiliation{Joint Institute for High Temperatures of the Russian Academy of Sciences, 
Izhorskaya 13, Bldg 2, Moscow 127412, Russia
}
%\author{M.~Bonitz}
%\author{Zh. Moldabekov}
%\affiliation{Institut f\"ur Theoretische Physik und Astrophysik, Christian-Albrechts-Universit{\"a}t zu Kiel, 
%Leibnizstrasse 15, 24098 Kiel, Germany}

\begin{abstract}
Path integral Monte Carlo simulations in the Wigner approach to quantum mechanics has been applied to calculate momentum and spin--resolved radial distribution functions of the strongly correlated soft--sphere quantum fermions.
The obtained spin--resolved radial distribution functions demonstrate arising triplet clusters of fermions, that is
the consequence of the interference of exchange and interparticle interactions.
The semiclassical analysis in the framework of the Bohr--Sommerfeld quantization condition applied to the potential 
of the mean force corresponding to the same--spin radial distribution functions allows to detect
exchange--correlation bound states in triplet clusters and to estimate corresponding averaged energy levels. 
The  obtained momentum distribution functions demonstrate the narrow sharp separated peaks corresponding to bound states and
disturbing the Maxwellian distribution. 
\end{abstract}
\maketitle

%05.30.Fk 	Fermion systems and electron gas
%05.70.Ce 	Thermodynamic functions and equations of state
%\begin{keywords}
%	exchange--correlation bound states, strongly correlated soft--sphere fermions, path integral Monte Carlo, sign problem
%\end{keywords}
%
\section{Introduction}\label{sec1} 

There are a number of simple model pair potentials for a strongly coupled system of particles, which are useful
in statistical mechanics and capable of grasping some physical properties of complex systems.
The examples include the soft- and hard-sphere fluids as well as the Lennard-Jones system.
The one component plasma is of great astrophysical importance being an excellent model for describing many features
of superdense, completely ionized matter typical for white dwarfs, the outer layers of neutron stars and possibly the interiors
of heavy planets \cite{luyten1971review,potekhin2010physics}.
Properties and structure of strongly correlated bosonic systems of hard- and soft-spheres are systematically
studied in the literature using different many-body approaches
\cite{boronat2000quantum,mazzanti2003energy,sese2020real,nozieres2018theory}. 
In such systems maxima in radial and momentum distribution functions and
the excitation spectrum are observed, which is a natural effect of the correlations when density increases.
The quantum interparticle correlations in triplets of particles are difficult to investigate
while the triplet correlations are important for any statistical analysis \cite{sese2020real}, as
they allow to formulate thermodynamic properties beyond the pairwise approach.

On the other hand, the theoretical studies of the strongly interacting particles obeying the Fermi--Dirac statistics
is a subject of general interest in many fields of physics.  
This work deals with the physical properties of a model system composed of strongly interacting soft-sphere
fermions at non-zero temperatures.
In the case of strong interparticle interaction perturbative methods cannot be applied, so direct computer simulations
have to be used.
At finite temperatures the most widespread numerical method in quantum statistics is a Monte--Carlo (MC) method
usually based on the representation of the quantum partition function in the form of path integrals
in the coordinate space of particles \cite{feynmanquantum,zamalin1977monte}. 
A direct computer simulation allows one to calculate a number of properties provided that the interaction potential is known.
For example, a path integral Monte-Carlo (PIMC) method is used to study thermodynamic properties of dense noble gases,
dense hydrogen, electron-hole and quark-gluon plasmas, etc.
\cite{EbelForFil,ForFilLarEbl,dornheim2018uniform,ceperley1995path,pollock1984simulation,singer1988path,filinov2022solution}.  

In this article we are going to use the PIMC method to study the properties
of strongly correlated soft--sphere fermions.
However the main difficulty of the PIMC method for Fermi systems is the ``fermionic sign problem''
arising due to the antisymmetrization of a fermion density matrix \cite{feynmanquantum} 
and resulting in thermodynamic quantities to be small differences of large numbers associated
with even and odd permutations. As a consequence, the statistical error in PIMC simulations grows
exponentially with the number of particles.
To overcome this issue a lot of approaches have been developed.
%(see, for example,  \cite{zamalin1977monte,EbelForFil,ForFilLarEbl,runeson2018quantum,dornheim2020attenuating}).  
%(\cite{zamalin1977monte,EbelForFil,ForFilLarEbl})
In \cite{ceperley1991fermion,ceperley1992path} to avoid the ``fermionic sign problem'', a restricted fixed--node path--integral 
Monte Carlo (RPIMC ) approach has been developed. In RPIMC only positive permutations are taken into account, so the accuracy of the results is unknown.
More consistent approaches are the permutation blocking path integral Monte Carlo (PB-PIMC) 
and the configuration path integral Monte Carlo (CPIMC) methods \cite{dornheim2018uniform}. 
In CPIMC the density matrix is presented as a path integral in the space of occupation numbers.
However it turns out that both methods also exhibit the ``sign problem'' worsening the accuracy of PIMC simulations.

An alternative approach based on the Wigner formulation of quantum mechanics in the phase space 
\cite{wigner1934interaction,Tatr} was used in \cite{larkin2017pauli,larkin2017peculiarities} to 
avoid the antisymmetrization of matrix elements and hence the ``sign problem''. This approach allows  
to realize the Pauli blocking of fermions and is able to calculate quantum momentum distribution functions
as well as transport properties \cite{EbelForFil,ForFilLarEbl}. 

We use the modified path integral representation of the Wigner function
and the MC approach (WPIMC)
to calculate the radial and momentum distribution functions of a soft--sphere fermionic system.
The WPIMC allows also to reduce the ``sign problem'' in contrast to a standard PIMC.
To overcome the ``sign problem'' the exchange interaction is expressed through a positive semidefinite
Gram determinant \cite{filinov2021monte}  in the expression for the density matrix.
%Due to the Gram determinant 
This article is the continuation of our publications
%\cite{filinov2021monte,filinov2022bound}. \cite{filinov2020uniform} \cite{filinov2015fermionic} 
on the improvements of the PIMC approach for strongly correlated systems of fermions and degenerate plasma media
\cite{zamalin1977monte,EbelForFil,ForFilLarEbl,filinov2015fermionic,filinov2020uniform,filinov2021monte,filinov2022bound}. 

We consider a 3D system of $N$ soft--spheres  obeying the Fermi--Dirac statistics
in the canonical ensemble at a finite temperature. In our approach the fermions interact through a quantum
pseudopotential corresponding to the soft--sphere potential
$\phi(x) =\epsilon (\sigma /x)^n$,
where $x$ is the interparticle distance, $\sigma$ characterizes the effective particle size,
$\epsilon$ sets the energy scale  and $n$ is a parameter determining the potential hardness. % (the potential softness, $s = n - 1$). 
The density of soft spheres is characterized by the parameter $r_s=a/\sigma$, defined as the ratio
of the mean distance between the particles $a=\left[3/(4 \pi \tilde{\rho}) \right]^{1/3}$ to $\sigma$  ($\tilde{\rho}$ is the number density).
For example, the results presented below have been obtained for the following physical parameters
used in \cite{filinov2022solution} for PIMC simulations of helium-3:
$\epsilon=26.7 {\rm K} $, $\sigma = 5.19 \, a_{\rm B}$ ($a_{\rm B}$ is the Bohr radius),
$m_a = 3.016$  is the soft--sphere mass in atomic units.
Other parameters are: $T=60 \, {\rm K}$, $r_s\approx 0.844$, so that $\lambda/\sigma \approx 0.48$, where $\lambda = \sqrt{2\pi\hbar^2\beta/m}$ is the thermal wavelength of a fermion, $\beta = 1/(k_B T)$ is the inverse temperature, $m$ is the mass of a fermion.
%($m_a=5.006E-27 kg $)   ($m_e=9.109E-31 kg$) ($1au=1.66E-27 kg$)
%Below in figures temperature will be given in units of the  quantum characteristic energy (explanation below )  
%$\tilde{\epsilon}={\rm Ha} \, m_e/m_a = 315 627\, {\rm K} \, m_e/m_a = 57.44\, {\rm K} $, 
%where ${\rm Ha}$ is the Hartree energy. 
%($m_a=5.006E-27 kg $)   ($m_e=9.109E-31 kg$) ($1au=1.66E-27 kg$)

In section II we consider the path integral description of quantum soft--sphere fermions.
In section III we derive a pseudopotential for the soft spheres accounting for the quantum effects in the interparticle interaction.
In section IV we present the results of our simulations.
The momentum distribution functions of the strongly coupled soft--sphere fermions 
calculated by WPIMC for different $n$ are discussed in subsection IV.A.
Subsection IV.B  deals with WPIMC spin--resolved radial distribution functions demonstrating
the short--range ordering of fermions (triplet fermion clusters) caused by the interference of the exchange
and interparticle interactions.  
In subsection IV.C the exchange--correlation bound states of fermion triplets and corresponding averaged energy levels
have been obtained from the Bohr--Sommerfeld quantization condition applied to the potential of the mean force
corresponding to the same--spin radial distribution functions.
In section V we summarize the basic results and discuss their physical meaning.

%%%%%%%%%%%%%%%%%%%%%%%%%%%%%
\section{Path integral representation of Wigner function}
Let us consider $N$ quantum soft--sphere fermions. The Hamiltonian of the system ${\hat H}={\hat K}+{\hat U}$
contains kinetic energy ${\hat K}$ and interaction energy ${\hat U}$ contributions 
taken as the sum of pair interactions $\phi(x)$.  
Since the operators of kinetic and potential energy do not commutate, the exact explicit 
analytical expression for the Wigner function is unknown
but can be formally constructed using a path integral approach~\cite{feynmanquantum,NormanZamalin,zamalin1977monte} based on the operator identity
$e^{-\beta {\hat H}}= e^{-\epsilon {\hat H}}\cdot
e^{-\epsilon {\hat H}}\dots  e^{-\epsilon {\hat H}}$,
where $\epsilon = \beta/M$ and $M$ is a positive integer number.
The Wigner function of the multiparticle system in the canonical ensemble
is defined as the Fourier transform of the off--diagonal
matrix element of the density matrix in the coordinate representation
\cite{wigner1934interaction,Tatr,feynmanquantum,ForFilLarEbl,zamalin1977monte,JAMP,LarkinFilinovCPP,Wiener,NormanZamalin}. 
The antisymmetrized Wigner function can be written in the form:
\begin{widetext}
\begin{multline}\label{permut}
	W(p,x;\beta) = \frac{1}{Z(N,V;\beta)N! \lambda^{3N}  }
	\sum_{\sigma}\sum_{P} (- 1)^{\kappa_{P}}  
	{\cal S}(\sigma, {\hat P} \sigma^\prime)
	\big|_{\sigma'=\sigma}\,
	\int {\rm d} \xi
	e^{i\langle \xi |p\rangle /\hbar} \langle x - \xi/2| e^{-\beta \hat{H} }|x + \xi/2\rangle \\
	\approx \frac{1}{Z(\beta)N! \lambda^{3N} }
	\sum_{\sigma}\sum_{P} (-1)^{\kappa_{P}}
	{\cal S}(\sigma, {\hat P} \sigma^\prime)
	\big|_{\sigma'=\sigma}\,
	 \int {\rm d} \xi
	e^{i\langle \xi |p\rangle /\hbar} \langle x - \xi/2| \prod_{m=0}^{M-1}
	e^{-\epsilon {\hat U_{m}}} e^{-\epsilon {\hat K_{m}}}|{\hat P}(x + \xi/2)\rangle.
\end{multline}
\end{widetext}
The partition function $Z$ for a given temperature $T$ and fixed volume $V$ is defined by the expression: 
\begin{equation}\label{q-def}
	Z(N,V;\beta) = \frac{1}{N! \lambda^{3N} } \sum_{\sigma}\int\limits_V {\rm d}x \,\rho(x, \sigma;\beta),
\end{equation}
where $\rho(x, \sigma;\beta)$ denotes the diagonal matrix
elements of the density operator ${\hat \rho} = e^{-\beta {\hat H}}$.
Here $x$ and $\sigma $ are  vector variables of the spatial coordinates and spin degrees of freedom
of particles, 
$\lambda=\sqrt{\frac{2\pi\hbar^2\beta}{m}}$ is the thermal wavelength.
The integral in Eq.~(\ref{q-def}) can be rewritten as:     
\begin{widetext}
\begin{multline}
	\sum_{\sigma} \int\limits {\rm d}x^{(0)}\,
	\rho(x^{(0)},\sigma;\beta) =
	%\nonumber\\ &&
	\int\limits  {\rm d}x^{(0)} \dots
	{\rm d}x^{(M-1)} \, \rho^{(1)} \dots \rho^{(M-1)}
	\sum_{\sigma}\sum_{P_e} (-1)^{\kappa_{P}} \,
	{\cal S}(\sigma, {\hat P} \sigma^\prime)\, %\times
	%\nonumber\\ &&
	{\hat P} \rho^{(M)}\big|_{x^{(M)}= x^{(0)}, \sigma'=\sigma}\, \\
	\approx {} \int{ {\rm d}x^{(0)} \dots{\rm d}x^{(M-1)}}
	\exp\Biggl\{-\sum_{m=0}^{M-1}\biggl[
	\pi \left|x^{(m)}-x^{(m+1)}\right|^2
	%/\check{\lambda}^2
	+\epsilon U(x^{(m)}) \biggr]\Biggr\}{\rm det}\|\Psi(x)\|, 
	\label{rho-pimc}
\end{multline}
\end{widetext}
where we imply that momentum and coordinate are dimensionless variables
$p \tilde{\lambda}/ \hbar$ and $x/ \tilde{\lambda}$
related to a temperature $\sim 1/\epsilon$   ($\tilde{\lambda}=\sqrt{2\pi\hbar\beta / (m M)}$).
Spin gives rise to the standard spin part of the density matrix 
${\cal S}(\sigma, {\hat P} \sigma^\prime)=\prod_{k=1}^N \delta(\sigma_k,\sigma_{Pk})$, 
($\delta(\sigma_k,\sigma_t)$ is the Kronecker symbol)  with exchange effects accounted for by the permutation
operator  $\hat P$  acting on coordinates of particles
$x^{(M)}$ and spin projections $\sigma'$. The  
sum is taken over all permutations with a parity $\kappa_{P}$.
In Eqs.~(\ref{permut}), (\ref{rho-pimc}) index $m=0,\dots , M-1$
labels the off--diagonal high--temperature density matrices 
$\rho^{(m)}\equiv \rho\left(x^{(m)},x^{(m+1)};\epsilon \right) =
\langle x^{(m)}|e^{-\epsilon {\hat H}}|x^{(m+1)}\rangle$. 
With the error of the order of $1/M^2$ each high--temperature factor can be presented in the form
$\langle x^{(m)}|e^{-\epsilon {\hat H}}|x^{(m+1)}\rangle \approx
\langle x^{(m)}|e^{-\epsilon {\hat U}}|x^{(m+1)}\rangle \rho^{(m)}_0$ 
with $  \rho^{(m)}_0=\langle x^{(m)}|e^{-\epsilon {\hat K}}|x^{(m+1)}\rangle$, 
arising from neglecting the commutator $\epsilon^2 \left[K,U\right] / 2$ and higher powers of $\epsilon$ terms.
In the limit $M\rightarrow \infty$ the error of the whole product of high temperature factors 
is equal to zero $(\propto 1/M)$ 
and we have an exact path integral representation of the Wigner and partition functions,  in which
each particle is represented by a trajectory consisting of a set of
$M$ coordinates (``beads''): 
$\tilde{x}\equiv\{x_{1}^{(0)}, \dots x_{1}^{(M-1)},
x_{2}^{(0)}\ldots x_{2}^{(M-1)}, \ldots x_{N}^{(M-1)}$.  
In the thermodynamic limit the main contribution in the sum over spin variables comes from the term related
to the equal numbers ($N/2$) of fermions with the same spin projection  \cite{EbelForFil,ForFilLarEbl}. 
The sum over permutations gives  the product of determinants: 
$ \mathrm{det} \|\Psi(x)\| = \mathrm{det} \bigl\|e^{-{\pi} \left|x_{k}^{(M)}-x_{t}^{(0)}\right|^2}\bigr\|_1^{N/2} 
\mathrm{det} \bigl\|e^{-{\pi} \left|x_{k}^{(M)}-x_{t}^{(0)}\right|^2}  \bigr\|_{(N/2+1)}^{N}$. 

In general the complex-valued integral over $\xi$ in the definition of the Wigner function (\ref{permut})
can not be calculated analytically and is inconvenient for Monte Carlo simulations. 
The second disadvantage is that Eqs.~(\ref{permut}), (\ref{rho-pimc})
contain the sign--altering  determinant ${\rm det}\|\Psi(x)\|$, which is the reason of the  
``sign problem'' worsening the accuracy of PIMC simulations.
To overcome these problems let us replace the variables of integration 
$x^{(m)}$ by $q^{(m)}$ 
%($q^{(0)}=q^{(M)}=0$) 
for any given permutation $P$ using the substitution
\cite{LarkinFilinovCPP,larkin2017peculiarities}:  
\begin{multline}
	x^{(m)} = (P x -x)\frac{m}{M}+x + q^{(m)} \\
	{} -\frac{ (M-m) \xi}{2M}+\frac{ m P\xi }{2M},
	\label{var}
\end{multline}
where $P$ is the matrix representing a permutation and equal to the unit matrix $E$ with appropriately transposed columns.
This replacement presents each trajectory as a sum of the ``straight line''
($(P x -x)\frac{m}{M} +x -\frac{ (M-m) \xi}{2M}+\frac{ m P\xi }{2M} $) 
and the deviation from it ($q^{(m)}$).
As a consequence the matrix elements of the density matrix can be rewritten  in the form of a path integral 
over \emph{``closed''}  trajectories $ \{q^{(0)}, \dots, q^{(M)} \}$ with $q^{(0)} =q^{(M)} =0 $ 
and after the integration over $\xi$ \cite{LarkinFilinovCPP,larkin2017peculiarities} 
and some additional transformations (see  \cite{filinov2020uniform, filinov2021monte} for details)  
the Wigner function can be written in the form containing the  Maxwell distribution with quantum corrections: 
\begin{widetext}
\begin{multline}
	W(p,x;\beta)
	\approx \frac{\tilde C(M)}{Z(\beta)N!}
	\int  {\rm d} q^{(1)} \dots {\rm d} q^{(M-1)} \exp\Bigl[ - \sum\limits_{m = 0}^{M-1}
	\biggl( \pi | \eta^{(m)}|^2 - \epsilon U\biggl(x + q^{(m)}\biggr)\biggr) \Bigr]
	\\
	\times \exp\Biggl\{\frac{M}{4 \pi}
	\left<  i p - \frac{\epsilon}{2}\sum\limits_{m = 0}^{M-1} \frac{ (M-2m) }{M}
	\frac{\partial  U(x + q^{(m)})}{\partial x} 
	\Bigg|  i p - \frac{\epsilon}{2}\sum\limits_{m = 0}^{M-1} \frac{ (M-2m) }{M}
	\frac{\partial  U(x + q^{(m)})}{\partial x}
	\right>
	% \Biggr|_{x + q^{(m)}}
	\Biggr\}
	\\
	%	\times {\rm det}\|\Phi(x)\|, \,
	\times \mathrm{det} \|\tilde{\phi}_{kt} \bigl\|_1^{N/2} \mathrm{det}\bigr\|\tilde{\phi}_{kt} \|_{(N/2+1)}^{N_e} , \,
	\label{rho-pimc44} 
\end{multline}
where 
\begin{equation}
	%	&&\sum_{\sigma}\sum_{P} %\sum_{P_p}
	%	(\pm 1)^{\kappa_{P}	} {\cal S}(\sigma, {\hat P} \sigma^\prime)
	%	\big|_{\sigma'=\sigma}\,
	%	%\nonumber    \\&&
	%	\exp\Biggl\{ - \pi \frac{|P x-x|^2}{M}-\epsilon \Delta U_{P} 
	%	\Biggr\} 
	%	\nonumber    \\&&
	%	%&&{\rm det}\|\Psi(x)\|=
	%	\approx 
	%	{\rm det}\|\Phi(x)\|=\det \|\tilde{\phi}_{kt} \bigl\|_1^{N/2}\times \mathrm{det}\bigr\|\tilde{\phi}_{kt} \|_{(N/2+1)}^{N_e}  
	%	\text{   and    }   
	%	\nonumber    \\&&
	\tilde{\phi}_{kt} =  \exp \{-{\pi} \left|r_{kt}\right|^2/M \}   %\times {} 
	%\nonumber    \\&&
	\exp \Biggl\{-\frac{1}{2}\sum\limits_{m = 0}^{M-1}\biggl( \epsilon \phi \Bigl(\Bigl|r_{tk}\frac{2m}{M} 
	+r_{kt} + q_{kt}^{(m)}\Bigr|  \Bigr) 
	- {} \epsilon \phi \Bigl(\Bigl|r_{kt} + q_{kt}^{(m)}\Bigr| \Bigr) \biggr)\Biggr\}, 
	\label{rho-pimcc} 
	\nonumber
\end{equation}
\end{widetext}
and $i$ is imaginary unit,  $\eta^{(m)} \equiv  q^{(m)} - q^{(m+1)}$, 
$r_{kt}\equiv (x_{k}-x_{t})$, $(k,t=1,\dots,N)$.   
The constant $\tilde{C}(M)$ is canceled in Monte Carlo calculations.

Let us stress that the approximate implementation of the Wigner function used in our simulations and specified
in Eq.~(\ref{rho-pimc44})
accounts for the contribution of all permutations in the form of the determinants
$\mathrm{det} \|\tilde{\phi}_{kt} \bigl\|_1^{N/2}\mathrm{det}\bigr\|\tilde{\phi}_{kt} \|_{(N/2+1)}^{N_e}$.  
Moreover, approximation (\ref{rho-pimc44}) have the correct limits to the cases of weakly and strongly degenerate
fermionic systems. Indeed, in the classical limit
the main contribution comes from the diagonal matrix elements due to the factor  $\exp \{-{\pi} \left|r_{kt}\right|^2/M \}$ and the differences of potential energies in the exponents
are equal to zero (identical permutation).  
At the same time, when the thermal wavelength is of the order of the average interparticle distance 
and the trajectories are highly entangled the term $r_{tk}\frac{2m}{M}$ in the potential energy
$ \phi \Bigl(\Bigl|r_{tk}\frac{2m}{M}  +r_{kt} + q_{kt}^{(m)}\Bigr|  \Bigr) $ can be omitted
and the differences of potential energies in the exponents tend to zero \cite{filinov2020uniform, filinov2021monte}.

\section{Quantum pseudopotential for soft--sphere fermions}\label{PsdPtt}

\begin{figure}[htp]
	\centering
	\includegraphics[width=0.8\columnwidth,clip=true]{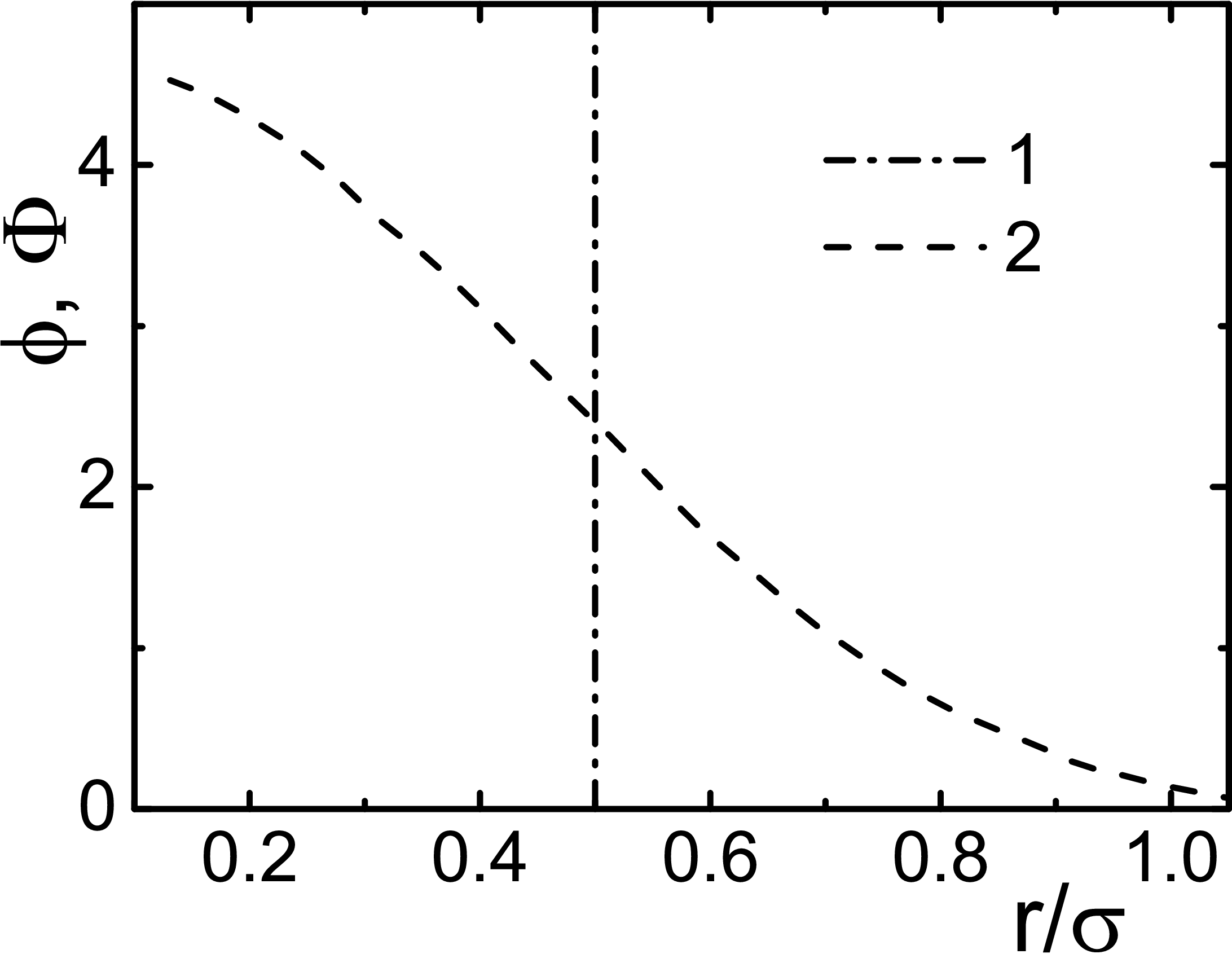}
	\includegraphics[width=0.9\columnwidth,clip=true]{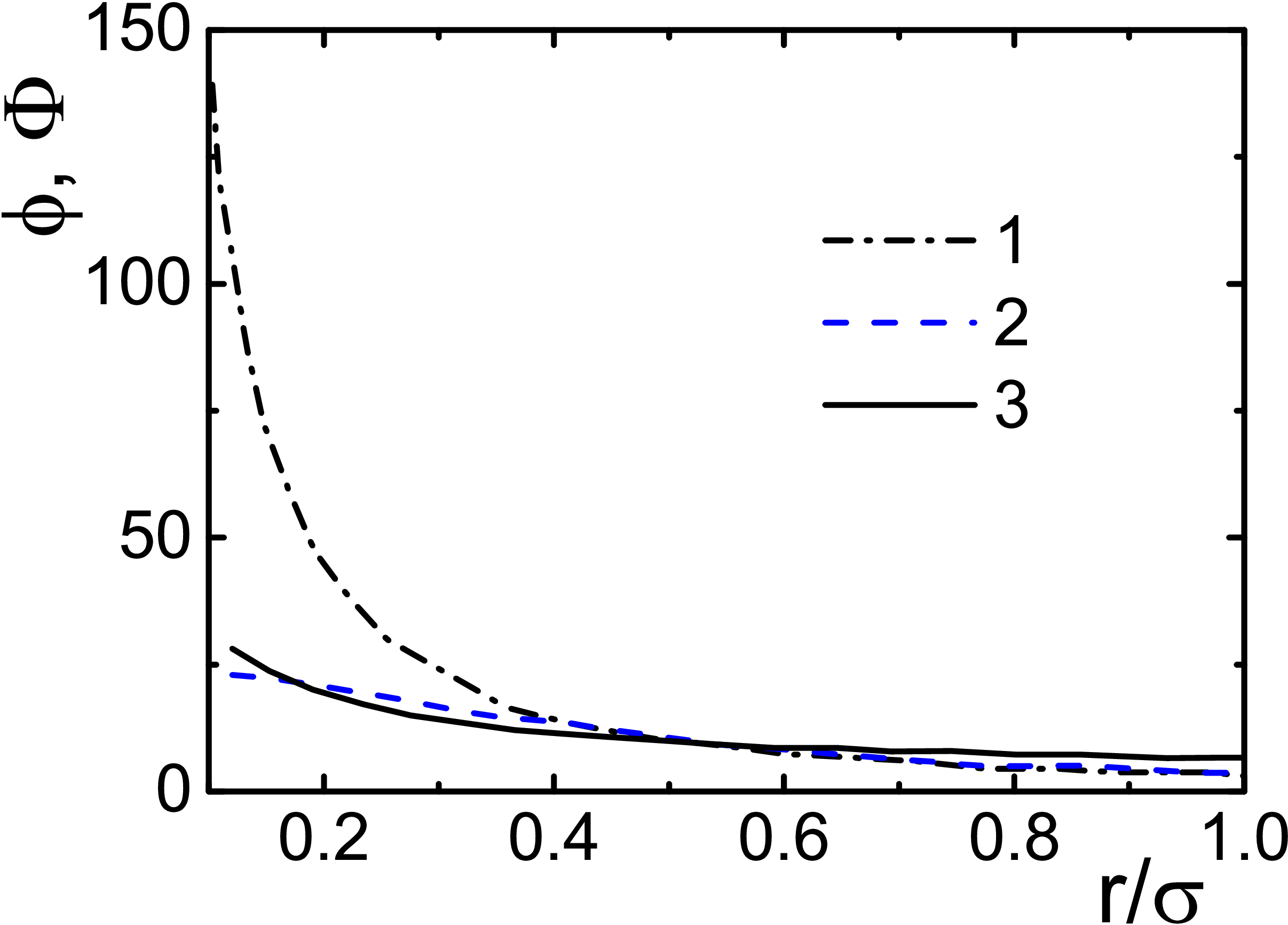}
	% \hspace{-3.2cm}
	%	\vskip 0.3cm
	%\hspace{.2cm}
	% \includegraphics[width=8.10cm,clip=true]{dskv}%
	\caption{(Color online) Potentials $\phi$ (lines 1) and corresponding pseudopotentials $\Phi$ (lines 2)
		defined by Eq.~(\ref{kelbg-d}) in conditional units.
		(Top panel) The bounded from above by the constant $h=10$ ``hard--sphere''  potential and the corresponding  $\Phi$.
		(Bottom panel) The soft--sphere potential ($n=5/3$) and   the corresponding $\Phi$
		approximated by the soft--sphere $\phi$ with $n \sim 2/3$ (line 3)).  
		\label{avrpt} 
	}
\end{figure}

As alternative the high--temperature density matrix $\rho^{(m)}=\langle
x^{(m)}|e^{-\epsilon {\hat H}}|x^{(m+1)}\rangle$ can be expressed as a
product of two--particle density matrices \cite{EbelForFil}
\begin{multline}\label{rho_ab}
\rho(x_{l},x'_{l}, x_{t}, x'_{t};\epsilon)
= \frac{1}{\tilde{\lambda}^6}
\exp\left[-\frac{\pi}{\tilde{\lambda}^2} |x_{l} - x'_{l}|^2\right] \\
%	\, \nonumber\\  	&&\times
{} \times \exp\left[-\frac{\pi}{\tilde{\lambda}^2} |x_{t} - x'_{t} |^2\right]
\exp[-\epsilon \Phi^{OD}_{lt}]\,.
\end{multline}
This formula  results from the factorization of the density matrix into the kinetic and potential parts,
$\rho \approx\rho_0^K\rho^{U}$. The off--diagonal density matrix element (\ref{rho_ab}) involves
an effective pair interaction by a pseudopotential, which can be expressed approximately via
its diagonal elements, $\Phi^{OD}_{lt}(x_{l},x'_{l},x_{t},x'_{t};\epsilon)
\approx [\Phi_{lt}(x_{l}-x_{t};\epsilon)+\Phi_{lt}(x'_{l}-x'_{t};\epsilon)]/2$.
To estimate $\Phi(x)$ for each high temperature density matrix  we use the well known semiclassical approximation \cite{feynmanquantum}:
\begin{eqnarray}
\epsilon \Phi(x;\epsilon) = \left(\frac{12}{{\tilde\lambda}_r^2}\right)^{3/2} \,
\int\limits_{-\infty}^{+\infty} \epsilon \phi(y+x) \exp \left[ {-\frac{12 \pi y^2}{\tilde\lambda_r^2}} \right] {\rm d^3}y,
\label{kelbg-d}
\end{eqnarray}
where $\tilde{\lambda}_r=\sqrt{2\pi \hbar^2 \epsilon / \tilde{m}}$,
$\tilde{m}=m/2$ is the reduced mass of a fermion.
Averaging  of the potential $\phi$  according to Eq.~(\ref{kelbg-d})  reduces the hardness of the soft--sphere potential. 
As an illustration, Fig.~\ref{avrpt} presents the bounded from above by the constant $h=10$ ``hard--sphere''  potential,
its pseudopotential, the soft--sphere potential $\phi$ for $n=5/3$, the corresponding  pseudopotentials $\Phi$
and its fitting soft--sphere approximation $\phi$ for $n \simeq 2/3$.

To derive a more accurate but more complicated pseudopotential $\Phi(x;\epsilon)$ for the potential $\phi(r)$  
%which accounts for quantum effects in the potential energy $U$, 
we have considered also the Kelbg functional \cite{demyanov2022derivation,Ke63} for the Fourier transform $v(t)$ of the potential $\phi(r)$. 
This transform can be found at $n < 3$ for the corresponding Yukawa--like potential $exp(-\kappa r)/r^n$ 
in the limit of ``zero screening''   ($\kappa \rightarrow 0 $:
\begin{equation}
	v(t) = \frac{4 \pi t^n \Gamma(2-n)\sin (n\pi/2 )}{t^3}.
	% 	\nonumber    \\&&
	%\label{kelbg-ff}
\end{equation}
where $\Gamma$ is the gamma function. The resulting quantum pseudopotential has the following form:
%%%\begin{multline}
\begin{multline}
	\Phi(x,\epsilon) =  \frac{\sqrt{\pi}}{8\pi^3} \times {}\\
	\int^{\infty}_0 v(t) \exp (-(\tilde{\lambda} t)^2/4) %{}\times
	\frac{ \sin (t r)\text{erfi}(\tilde{\lambda} t/2)}{(t r) \tilde{\lambda}t} 4 \pi t^2 {\rm d}t,
\end{multline}
%%%\end{multline}
where $\text{erfi}(z)= i\text{efr}(iz)$, $\text{efr}(z)$  is the error function, $r=|x|$ \cite{demyanov2022derivation}. 
This pseudopotential is finite at zero distance and decreases according to the power law with increasing distance. 
  
For more accurate accounting for quantum effects the ``potential energy''  $U(x^{(m)},x^{(m+1)})$ 
in (\ref{permut}) and (\ref{rho-pimc}) has to be taken as the sum of pair interactions given  
by $\Phi^{OD}$ with $\Phi(x;\epsilon)$. 
However, if the effective hardness of the pseudopotential $\Phi$ is less than $3$  the corresponding energy $\sum_{m=0}^{M-1} \epsilon U(x^{(m)})$
may be divergent in the thermodynamic limit.   
% \cite{hansen1973statistical}.
%\cite{sese2020real}
%%%%%%%%%\cite{sese2020real}  %%%%%%%%%%%%%%%%%\cite{hansen1973statistical}.
%, which conserve the thermodynamic stability of the fermion syaytem \cite{ruelle1999statistical} but
%and leads to energy divergence \cite{hansen1973statistical}     \cite{sese2020real}.
To overcome this deficiency let us modify the pseudopotential  $\Phi$
according to the transformation considered in \cite{hansen1973statistical}:
\begin{multline}
\tilde{\Phi}(x;\epsilon) = \epsilon  [ \Phi(x;\epsilon) -
\frac{1}{V}\int\limits_{V}{\rm d^3} y \Phi (x+y;\epsilon) ] \\
{} =\epsilon \int\limits {\rm d^3} y \Phi(x+y;\epsilon) \left(\delta (y) - \frac{1}{V}\right).
\label{kelbg-d2}
\end{multline}
Here the uniformly ``charged'' background is introduced to compensate the possible divergence of $U(x^{(m)})$
like in the one-component Coulomb plasma. 
 
Systems of 100, 200 and 300 particles represented by twenty and forty ``beads'' interacting with 
the quantum pseudopotential $\tilde{\Phi}(x;\epsilon)$ given by approximations Eq.~(\ref{kelbg-d}) 
as well as  Eq.~(\ref{kelbg-d2}) have been considered  in the basic Monte Carlo cell with periodic boundary conditions.    
The WPIMC configurations in the range of the $10^6 - 3\times 10^6$ of the Markovian chain have been generated  
to calculate distribution functions. 
So we have checked the convergence of the calculated distribution functions with increasing number 
of particles represented by the increasing number of beads in this range of parameters. 

Let us note that the pseudopotential corresponding to the Coulomb potential with hardness $n=1$ was often used in PIMC simulations 
of one-- and two--component plasma media in \cite{Ke63,kelbg,afilinov-etal.04pre,ebeling_sccs05,KTR94,EbelForFil,ForFilLarEbl} 
with good agreement with available in literature data. We used here the potentials with a hardness a bit more and less than unity. 
In the considered here range of hardnesses the convergence of the distribution functions was tested and  
it turns out that 300 particles represented by 20 beads is enough to reach the convergence.  

%%%%%%%%%%%%%%%%%%%%%%%%%%%%%%%%%%%%%%%%   

\section{Simulation results}\label{simulations}  

In PIMC simulations some trajectories presenting fermions and starting in the basic Monte Carlo cell with periodic boundary conditions  
can ``cross'' a cell boundary. In this case the choice between the  ``basic cell bead'' and  its periodic image
in the calculation of the interparticle interaction $U(x^{(m)},x^{(m+1)})$ becomes ambiguous that is discussed in detail
in \cite{EbelForFil,ForFilLarEbl}.  Here this problem prevents making use of the Ewald technique for 
the calculation of the pseudopotential energy $U(x^{(m)},x^{(m+1)})$ in (\ref{permut}) and (\ref{rho-pimc}). 
%\cite{EbelForFil,ForFilLarEbl}.   
The second problem is the slow decay of the  pseudopotential $\tilde{\Phi}$. 

Due to these problems we present only radial and momentum distribution functions, which demonstrate faster convergence with increasing 
number of particles in a Monte Carlo cell %($N \ge 100$)
in comparison with thermodynamic quantities \cite{filinov2020uniform}. 
The pair distribution function \cite{kirkwood1935statistical,fisher1964statistical} 
and momentum distribution function (MDF) can be written in the form: 
%by the following expressions: 
\begin{eqnarray}  \label{gab-rho}
&&g_{ab}(r) = \int\limits_{V} \frac{dpdx}{(2\pi)^{6N}} \,\delta(|x_{1,a}-x_{1,b}|-r)\,
W(p,x;\beta),
\nonumber\\ &&
w_a(|p|) = \int\limits_{V}\frac{dpdx}{(2\pi)^{6N}} \,\delta(|p_{1,a}|-|p|)\,
W(p,x;\beta),
\end{eqnarray} 
where $\delta$ is the delta function, $a$ and $b$ labels the spin value of the fermion.
The pair distribution function $g_{ab}$ give the probability density to find a pair of particles of types $a$ and $b$
at a certain distance $r$ from each other and depends only on the difference of coordinates because of the
translational invariance of the system. 
In a noninteracting classical system, $g_{ab}  \equiv 1 $, whereas interaction
and quantum statistics result in a redistribution of the particles. 
The momentum distribution function $w_a(|p|)$ gives a probability density for particle of type $a$ to have a momentum $p$.

%Here the RDF can be written in the form \cite{kirkwood1935statistical,fisher1964statistical},
%\begin{eqnarray}
%g(r) = g(R_1,R_2) =
%%\nonumber\\
%\sum_{\sigma} \int\limits_{V} \, {\rm d} x  
%\,\delta(R_1-x_{1})\, \delta(R_2-x_{2}) \rho(x, \sigma;\beta)/Z(N, V, \beta),
%\label{gab-rho}
%%\label{q-def}
%\end{eqnarray}
%where $r=R_1-R_2$. 
%In an isotropic system an RDF depends only on the relative distance between particles $r=|R_1- R_2|$. 
% and is proportional to the probability density to find a pair of particles at a distance  $r$ from each other. 
%In a non-interacting classical system $g_{ab}(r)\equiv 1$, whereas interparticle interaction and quantum statistics result in
%a redistribution of particles. The product $r^2 g_{ab}(r)$ is proportional (up to a constant factor) to
%the probability to find a pair of particles at a distance $r$ from each other. 

\subsection{Momentum distribution functions}  

\begin{figure}[htp] 
	\centering
	\includegraphics[width=0.8\columnwidth,clip=true]{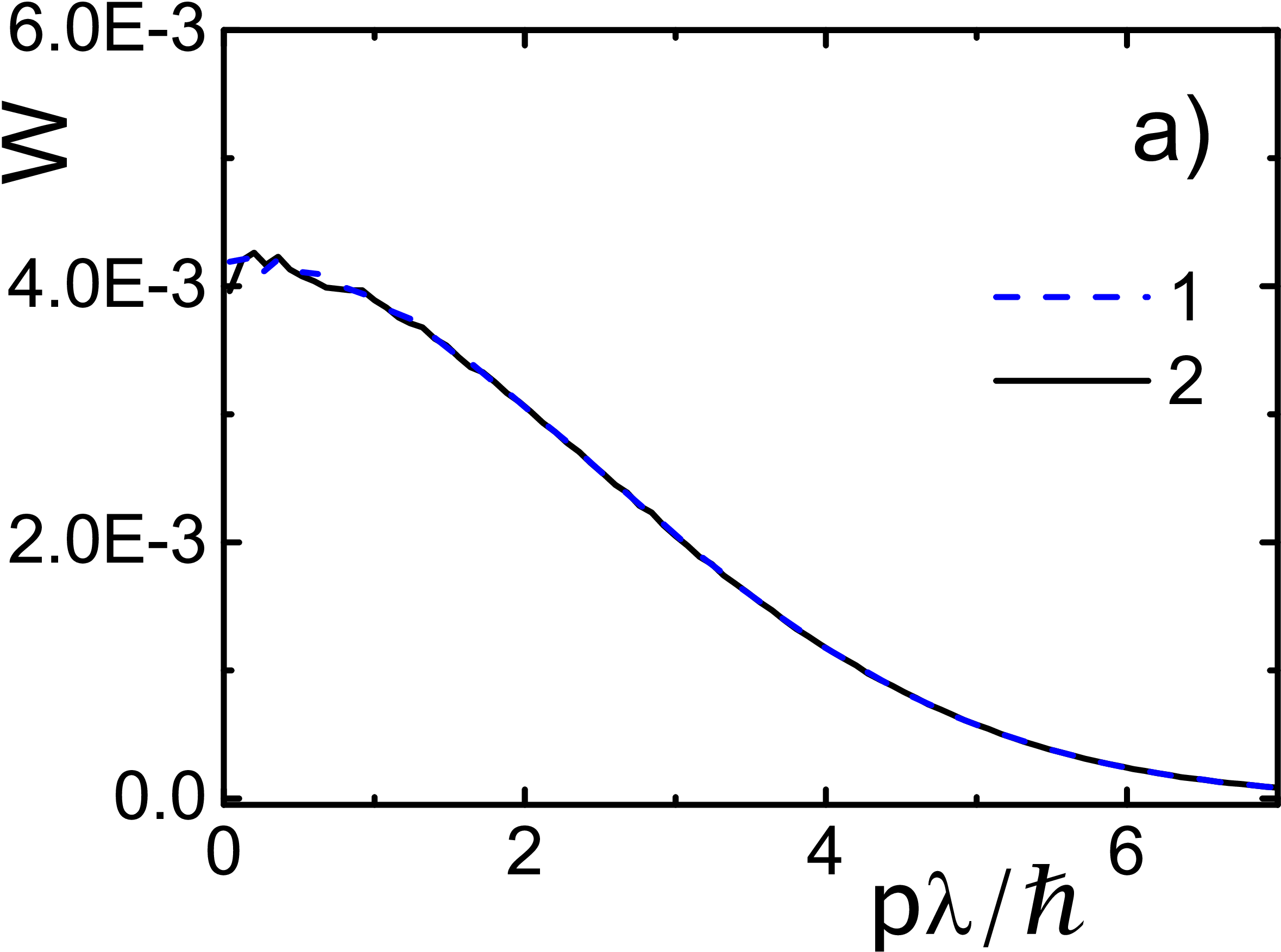}
	% \hspace{-3.2cm} 
	\includegraphics[width=0.8\columnwidth,clip=true]{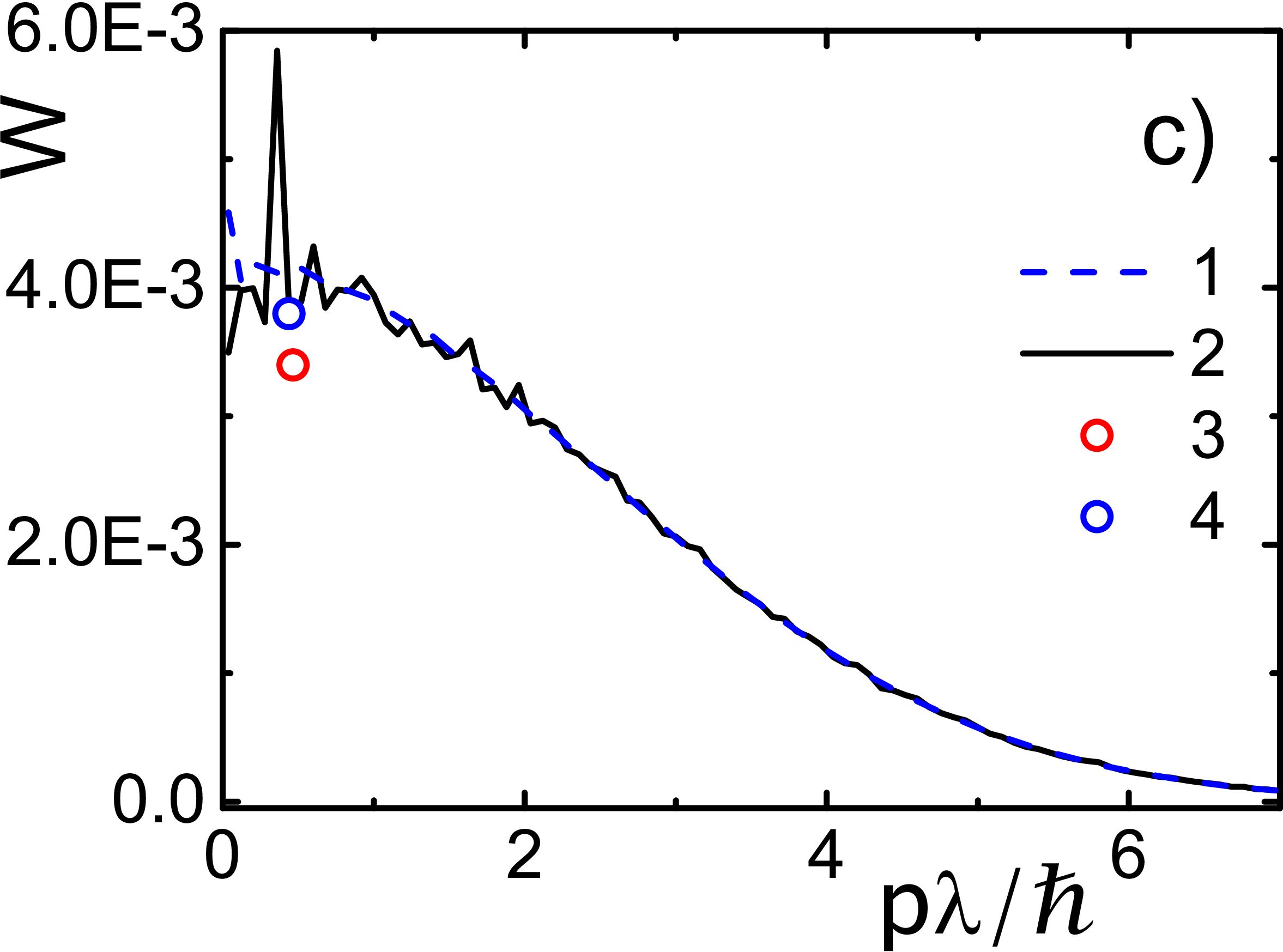}
%	\vskip 0.3cm
	%   \includegraphics[width=7.1cm,clip=true]{int70}  
	\includegraphics[width=0.8\columnwidth,clip=true]{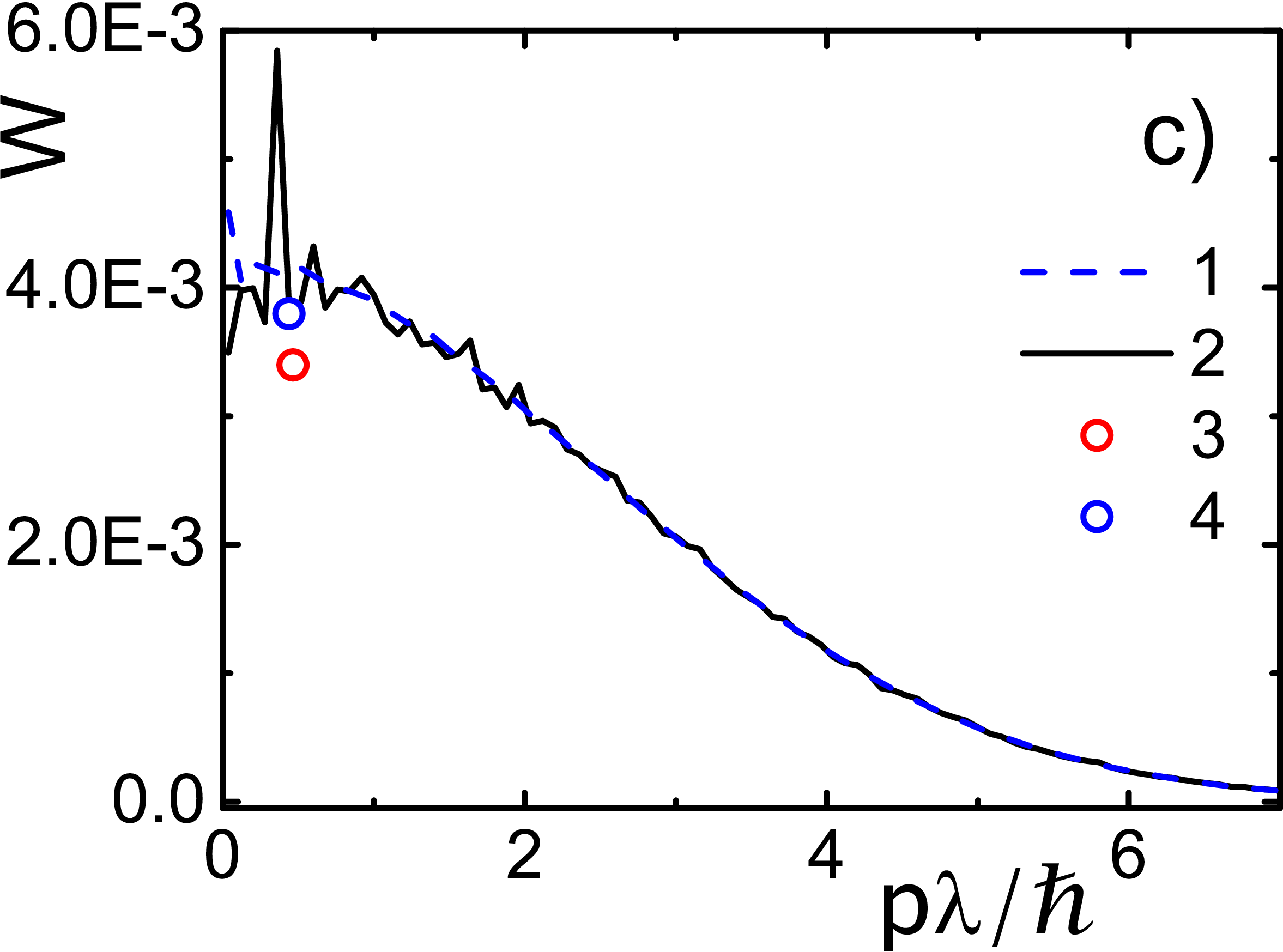}
	\includegraphics[width=0.8\columnwidth,clip=true]{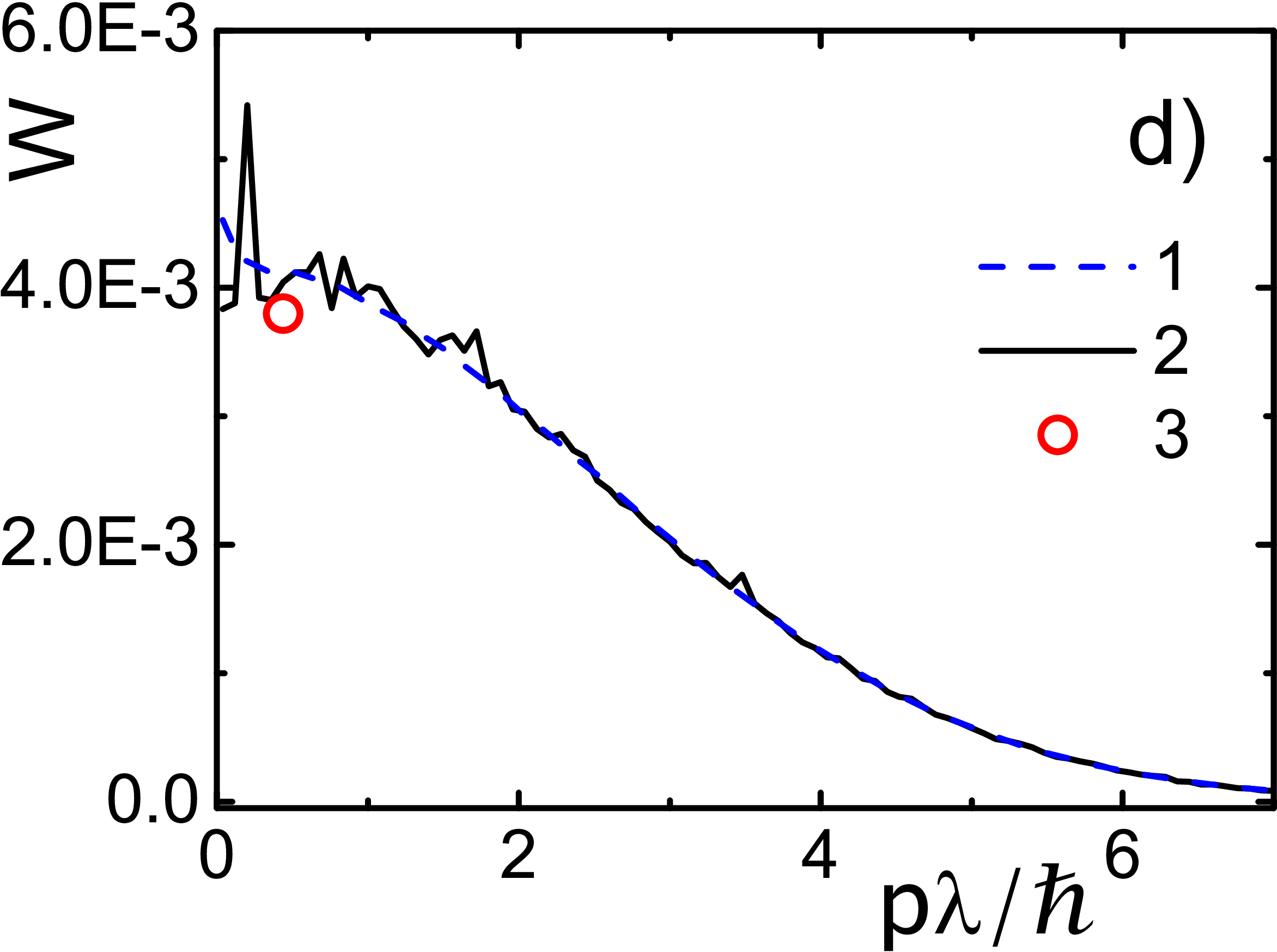}
	%\hspace{.2cm}
	% \includegraphics[width=8.10cm,clip=true]{dskv}% 
	\caption{(Color online) 
		WPIMC MDF $W(p\lambda /\hbar)$ for soft--sphere potentials with
		$n=0.2$---a); $n=0.6$---b); $n=1.0$---c); $n=1.4$---d).
		Lines: 1---the Maxwell MDF; 2---the WPIMC MDF for the same spin projections.
		Circles: 3, 4---momenta corresponding to the energy level with  $n_r=0$ and $L=0$, $L=1$ (see below).
		Small irregular oscillations of the MDFs indicate Monte-Carlo statistical error.
		\label{pwp} 
	}
\end{figure}

Figure~\ref{pwp} presents the Maxwell momentum distribution function ($\sim  \exp (-(p\lambda/\hbar)^2/4 \pi$)  (line $1$)
and the WPIMC calculations of the momentum distribution functions (MDFs) for the potentials with
$n=0.2$ (panel a)),  $n=0.6$ (panel b)), $n=1.0$ (panel c)), $n=1.7$ (panel d))
for the mentioned above fixed density ($r_s\approx 0.844$) and temperature ($T=60 \, {\rm K}$).
All MDFs are normalized to unity.
For a small hardness ($n=0.2$) of the pseudopotential as well as at large momentum
the  Maxwell distributions practically coincide with the WPIMC MDFs (all lines 2).

At the same time at a larger hardness ($n \geq 0.6$) and at smaller momentum lines 2 show the narrow separated  peaks
disturbing the Maxwell distribution. 
The physical meaning of these peaks as well as the circles  $3$ and $4$ are discussed below.

The convergence and statistical error of distribution functions with increasing number of the steps in the Monte Carlo runs
have been tested for increasing number of particles and number of beads at different softnesses of pseudopotentials. 
The width of the narrow sharp peaks in numerical distribution functions depends not only on the physical properties of the system 
but is also affected by the smallness of the discrete interval in the corresponding calculated histogram. 
With decreasing histogram interval the statistical errors increases due to worsening of statistics in each interval, 
so the compromise between reasonable value of  discrete interval and the statistical errors have to be achieved. 
In our simulations the values of statistical errors  are of the order of small random oscillations of the distribution functions,
which are several times less than regular peaks.

\subsection{Radial distribution functions and the potential of mean force}  

To explain the existence and physical meaning of the individual separated sharp high peaks on the MDFs
let us consider a typical random pseudopotential field created by  $\Phi(x;\epsilon)$, radial distribution functions (RDF)
and the corresponding potentials of mean interparticle force.

\begin{figure}[htp]
	\centering
	\includegraphics[width=0.8\columnwidth,clip=true]{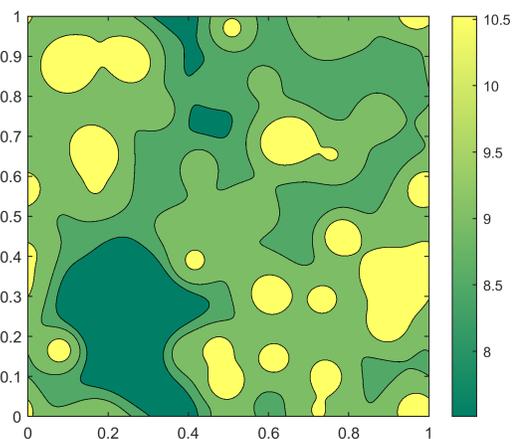}
	%	\vskip 0.3cm
	%\hspace{.2cm}
	\caption{(Color online)Typical random pseudopotential field in a cross section of the Monte Carlo cell for
		the particles interacting with $\Phi(x;\epsilon)$  at $n=2/3$.
		The vertical scale bar demonstrates the pseudopotential field variation in arbitrary units.
		\label{sec}
	}
\end{figure}

Fig.~\ref{sec} shows a typical random pseudopotential field created by  $\Phi(x;\epsilon)$ in a cross-section of the Monte Carlo cell. 
The vertical scale bar allows to estimate the pseudopotential field variation. 
% and average uniform background ``compensating  charge''.
As we shall see below from the consideration of RDFs and potentials of mean force
the possible formation of cavities separated by barriers
%in the repulsive soft--sphere--like systems of particles
is the physical reason of arising bound states for a triplet of fermions.

The potential of mean force  (PMF) $w^{(k)}$  \cite{kirkwood1935statistical} of a classical $N$ particle system is
defined up to an arbitrary constant as:  
\begin{eqnarray}
	-\nabla _{j}w^{(k)}\,=\,{\frac {\int e^{-\beta U}(-\nabla _{j}U)dq_{k+1}\dots dq_{N}}{\int e^{-\beta U}dq_{k+1}\dots dq_{N}}},
\end{eqnarray}
where $j=1,2,\dots, k$.
Above, $-\nabla _{j}w^{(k)} $ is the averaged force, i.e. the ``mean force'' acting on a particle $j$.
For $k=2$ the $w^{(2)}$ is related to the RDF of the system as
\begin{eqnarray}
	g(r)=e^{{-\beta w^{{(2)}}(r)}}. 
\end{eqnarray} 
Let us note that an RDF can be expressed as a virial expansion which in the low-density limit is given by
the formula \cite{fisher1964statistical,zelener1981perturbation}:
\begin{eqnarray}
	g(r)=e^{-\beta \phi(r)}.
\end{eqnarray} 
So the PMF $w^{{(2)}}(r)$ obtained from simulations determine to some extent
the effective interparticle interaction and corresponding physical properties.  

\begin{figure}[htp] 
	\centering
	\includegraphics[width=0.8\columnwidth,clip=true]{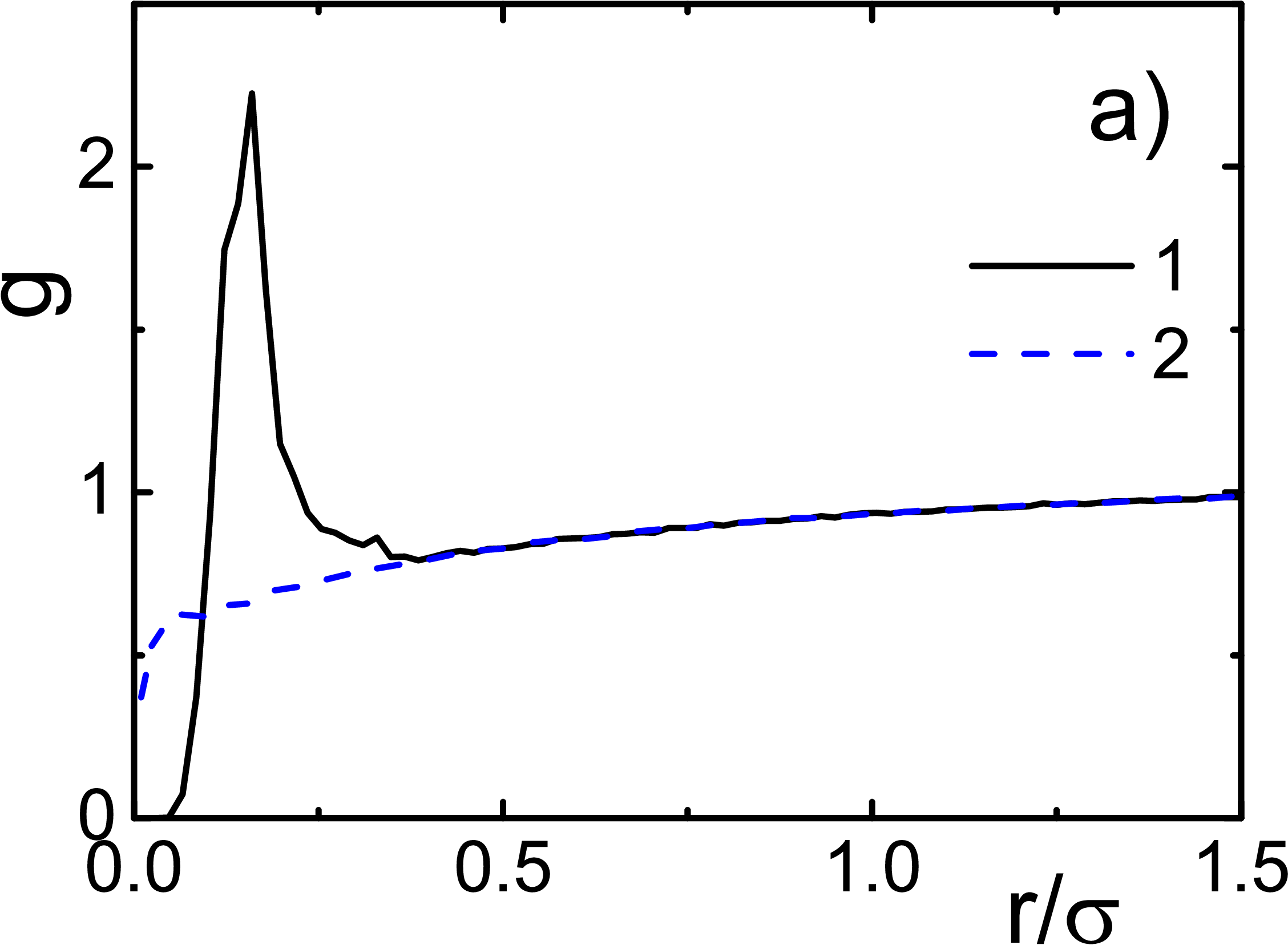}
	% \hspace{-3.2cm} 
	\includegraphics[width=0.8\columnwidth,clip=true]{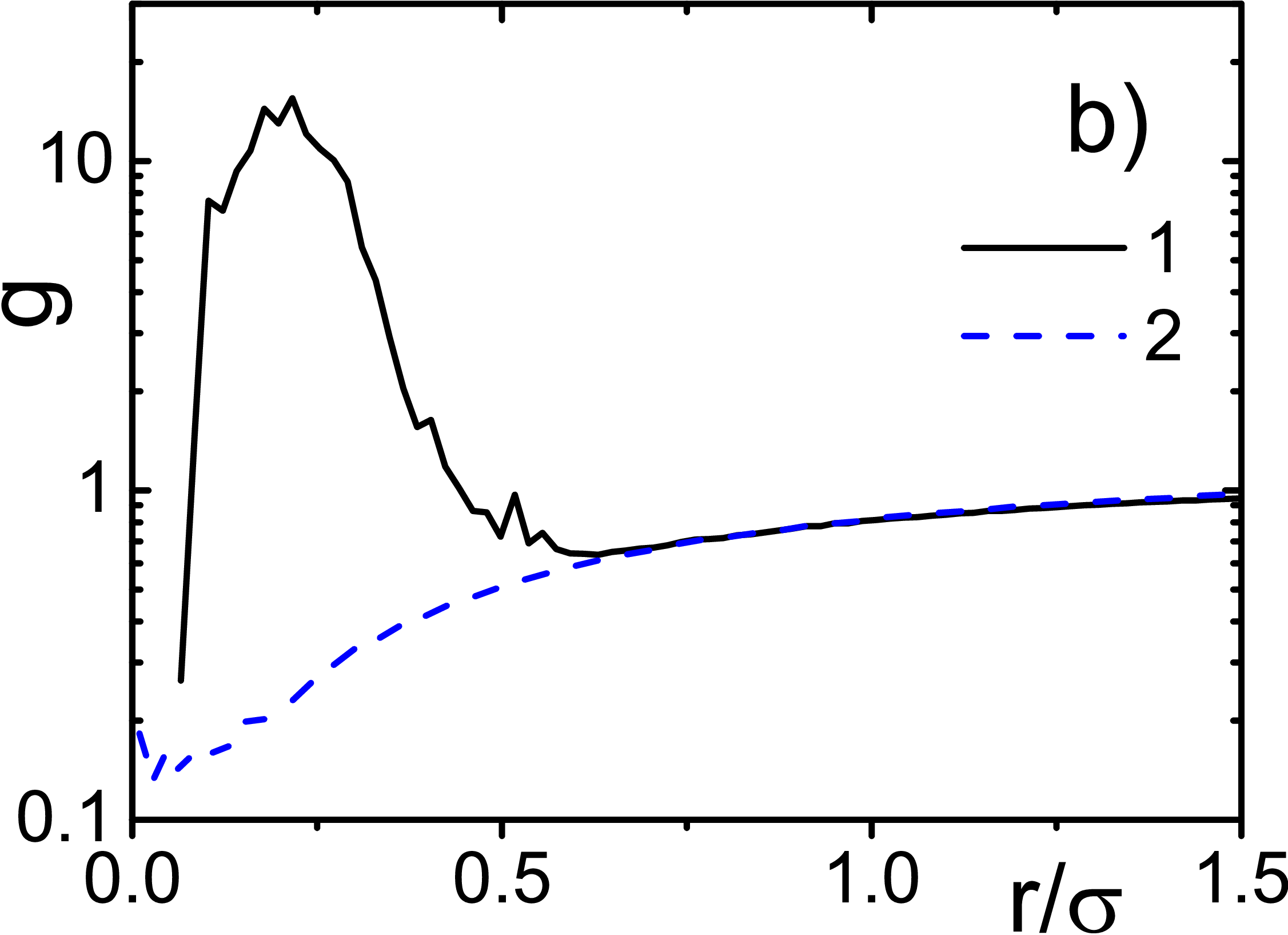}
%	\vskip 0.3cm
	\includegraphics[width=0.8\columnwidth,clip=true]{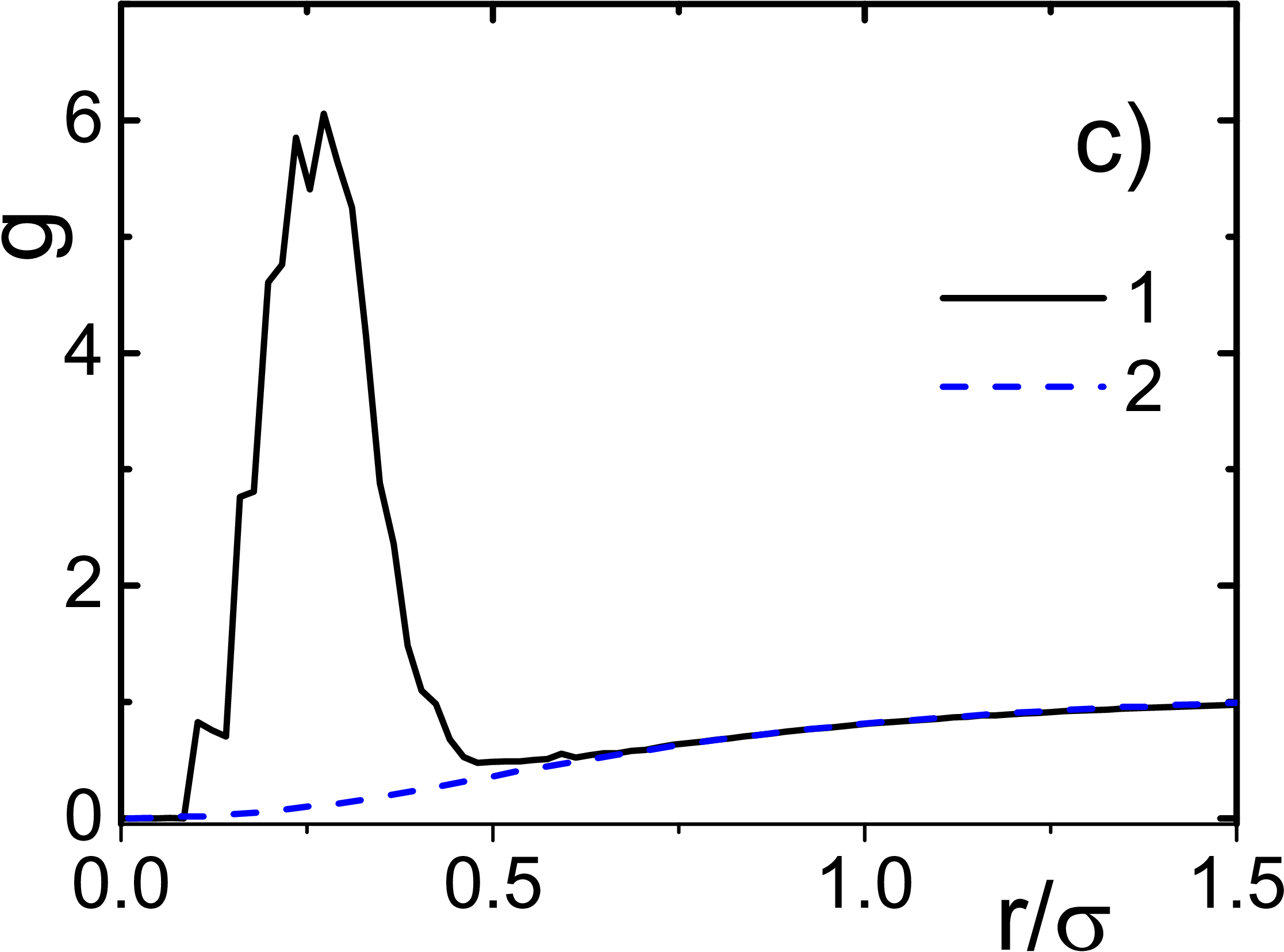}
	\includegraphics[width=0.8\columnwidth,clip=true]{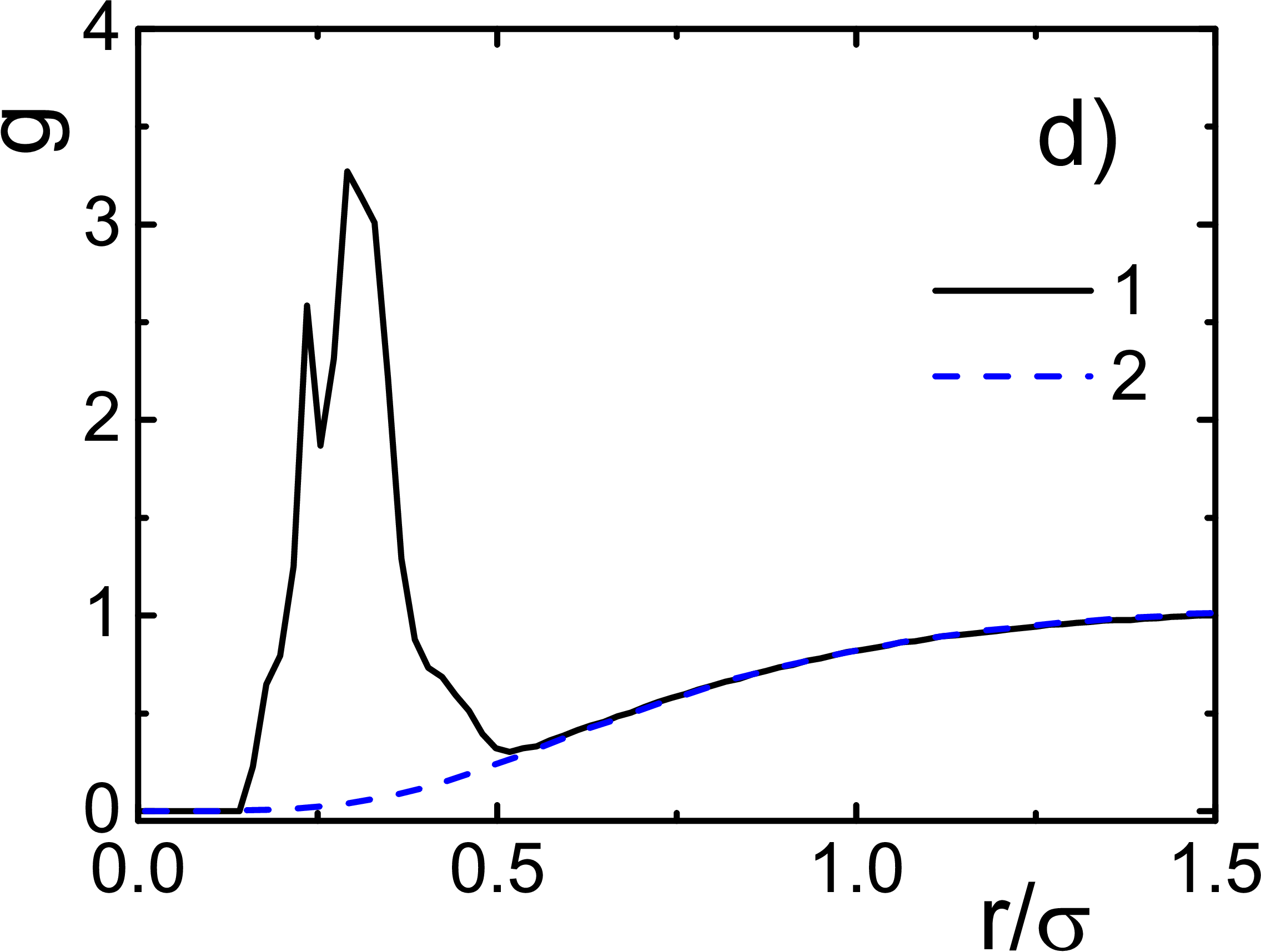}
	%\hspace{.2cm}
	\caption{(Color online)
		The RDFs  $g(r/\sigma)$ at
		$n=0.2$---a); $n=0.6$---b); $n=1.0$---c); $n=1.4$---d).
		%Panels: a) -- $n=0.2$; b) -- $n=0.6$; c) -- $n=1.0$; d) -- $n=1.4$. \\
		Lines: 1---RDF for the same spin projections; 2---RDF for the opposite spin projections.
		Small oscillations of the RDFs indicate the Monte-Carlo statistical error.  
		\label{drp} 
	}
\end{figure}

Figure~\ref{drp} presents the results of our WPIMC calculations for the RDFs with the same and opposite spin projections
for a fixed density and temperature but at different hardnesses of the soft--sphere pseudopotential.
Let us discuss the difference revealed between the RDFs with the same and opposite spin projections.
At small interparticle distances all RDFs tend to zero due to the repulsion nature of the soft--sphere potential.
Additional contribution to the repulsion of fermions with the same spin projection at distances of the order of the thermal wavelength are caused by the Fermi statistics effect described by the exchange determinant in (\ref{rho-pimc44}),
which accounts for the interference effects of the exchange and interparticle interactions. 
This additional repulsion leads to the formation of cavities (usually called exchange--correlation holes)
for fermions with the same spin projection and results in the formation of high peaks on the corresponding RDF
due to the strong excluded volume effect \cite{barker1972theories}.
The RDFs for fermions with the same spin projection show that the characteristic  ``size'' of an exchange--correlation cavity 
with corresponding peaks is of the order of the quantum soft--sphere thermal wavelength ($\lambda/\sigma\sim 0.48$), 
which is here less than the average interparticle distance ($r_s=0.844$). 
Let us stress that the strong excluded volume effect was 
also observed in  the classical systems of  repulsive particles (system of the hard spheres) seventy years ago 
in \cite{kirkwood1950radial} and was derived analytically for 1D case in \cite{fisher1964statistical}. 
%An RDF for the opposite spin projections shows that a fermion with the opposite spin to the discussed above pair of fermions 
%with the same spin can be in a cavity.
  
With increasing hardness of the potential the height and width of the RDF's peak are changing non-monotonically
reaching their maximal values at $n=0.6$.
The changes in the depth and width influence the PMF.
Let us stress that for opposite spin fermions the interparticle interaction is not enough to form any peaks
on the RDF (compare lines 1 and 2 in Figure~\ref{drp}a).
At larger interparticle distance the RDFs decay monotonically to unity due to the short--range repulsion of the potential.

\subsection{Exchange--correlation bound states} 

%$\Delta \omega_D / \omega \sim 2\sqrt{2} v_T/c $  \\
%$v_T=\sqrt{2k_BT/m_a}$
%$\Delta \omega_P / \omega \sim 7.2 {(C_{n}/\hbar)}^{(2/5)} (\sqrt{4k_BT/m_a})^{(3/10)}/\omega $ \\
%$\phi(r) \sim C_{n}/r^n$\\
Let us note that the peak on the RDFs for fermions with the same spin projection points out to the increase in the probability
to find a fermion between the two ones with the same spin projection at a distance a bit more than the peak position.
This third fermion can be considered as located in the well corresponding to the potential of mean force.
So these three fermions with the same spin projection can form a three--fermion cluster (TFC) or a triplet. The semiclassical approach
is used below to consider the possibility of bound states formation in such a TFC. Analogous exchange--correlation clusters
of electrons have been discussed in \cite{Weisskopf:PR:1939,lowd,filinov2021monte} 
for a plasma medium and have been identified with three--particle exchange--correlation excitons
(see also the collection of theory projects addressing long-standing questions in physics
by Prof. Emeritus Franz J. Himpsel \cite{Unsolved}). 

The semiclassical approach, which is known to be very effective for 
many problems of quantum mechanics and mathematical physics, is used below
to analyze the possibility of arising bound states in a TFC.
For this purpose let us use the Bohr--Sommerfeld condition \cite{nikiforov2005quantum} for a particle
in a spherically symmetric field $w^{(2)}$. 
According to the definition the PMF is determined up to an arbitrary constant,   %(mentioned above)
so to agree with the virial expansion  at low density
we have to assume here that $w^{(2)}(r) = 0$ in the limit $r\rightarrow \infty$.
So the Schr\"odinger equation for the radial part $R(r)$ of the wave function in atomic units
%in spherical coordinates 
looks like:   
\begin{multline}
	-\frac{1}{2} R^{\prime \prime}+ \frac{L(L+1)}{2r^2}R  -\left(\frac{k_B T}{{\rm Ha}}\right) \frac{m_a}{m_e} \ln (g(r))R \\
	{}=\left(\frac{E_L}{{\rm Ha}}\right) \frac{m_a}{m_e} R,
\end{multline}
where  $E_L$ is the energy level, $L=0,1,2, \dots$ is the orbital quantum number \cite{nikiforov2005quantum},
${\rm Ha} \, m_e /(m_a k_B)\approx 57.44 \, {\rm K}$.
In the semiclassical approximation for a particle in the PMF  field  $w^{(2)}$, the Bohr--Sommerfeld condition takes the form: 
\begin{eqnarray}\label{BrnSm}
	\int\limits_{ r_1(E)}^{ r_2(E)}p(r) {\rm d} r = \pi 
	\left(n_r +\frac{1}{2} \right),
	\label{BS} 
\end{eqnarray} 
where 
%we assumed that 
\begin{eqnarray}
	p(r)=\sqrt{2\left[\tilde{E}  + \left(\frac{k_B T}{{\rm Ha}}\right) \frac{m_a}{m_e} \ln (g(r)) \right]-\frac{(L+\frac{1}{2})^2}{r^2} },
	\label{pr}
\end{eqnarray} 
where $\tilde{E}=\big(\frac{E}{{\rm Ha}}\big) \frac{m_a}{m_e} $,  $n_r$ is the number of zeros of $R(r)$ \cite{nikiforov2005quantum}
and for any energy $E \ge w^{(2)}_{min}$ there are only two turning points $r_1$ and $r_2$. 
  
\begin{figure}[htp] 	
	\centering
	% \hspace{-3.2cm} 
	\includegraphics[width=0.8\columnwidth,clip=true]{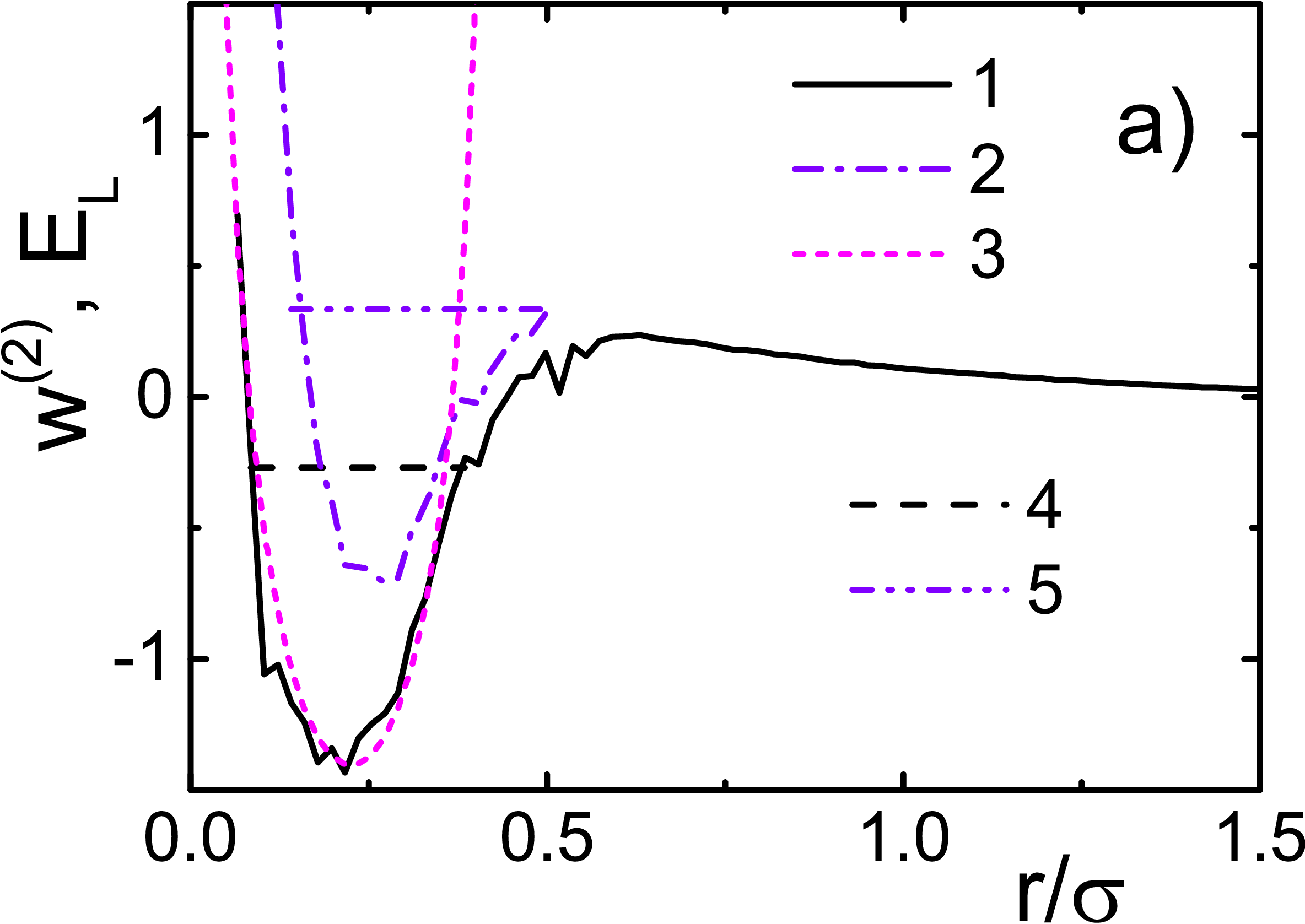}
	%	\vskip 0.3cm
	\includegraphics[width=0.8\columnwidth,clip=true]{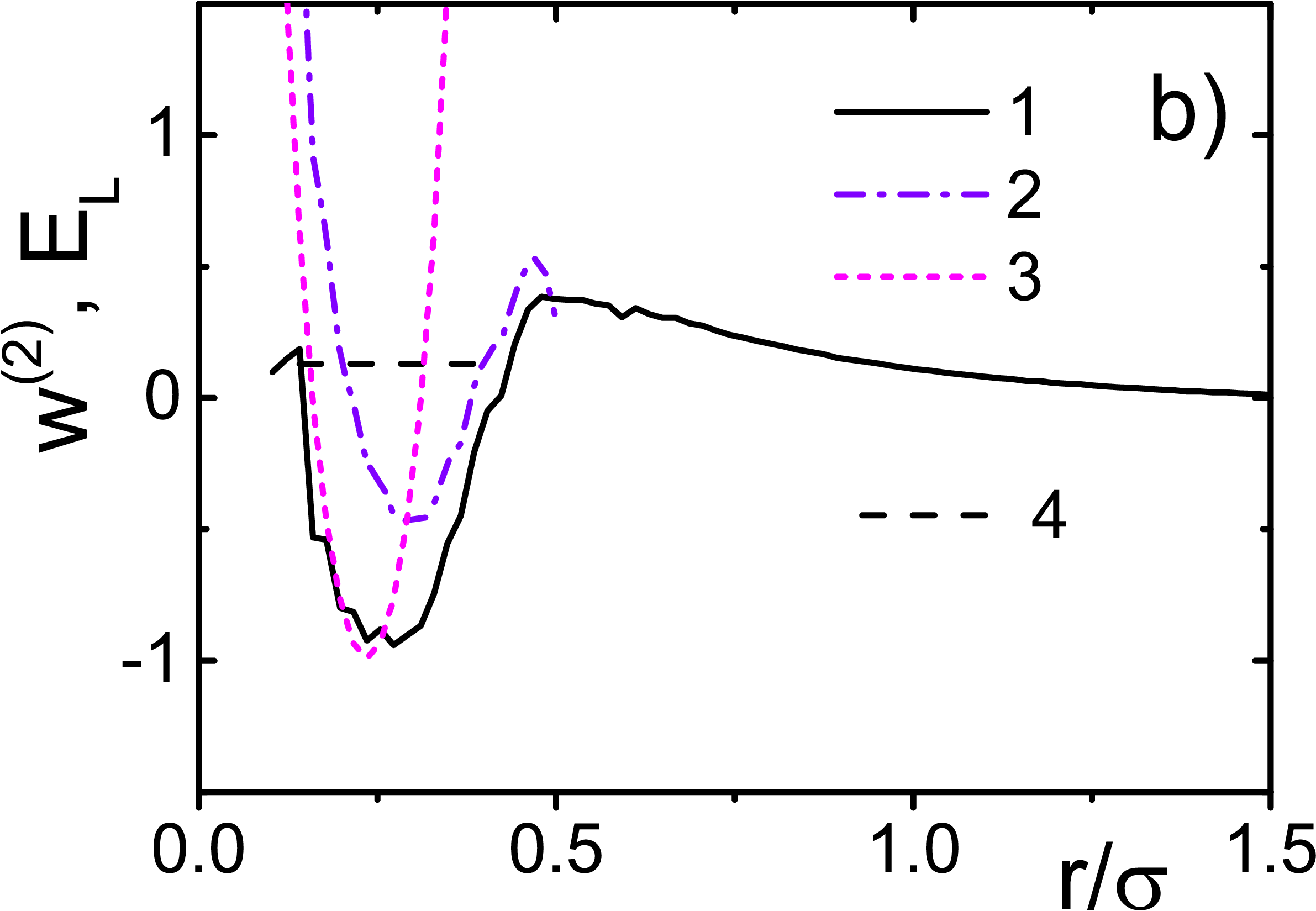}
	\includegraphics[width=0.8\columnwidth,clip=true]{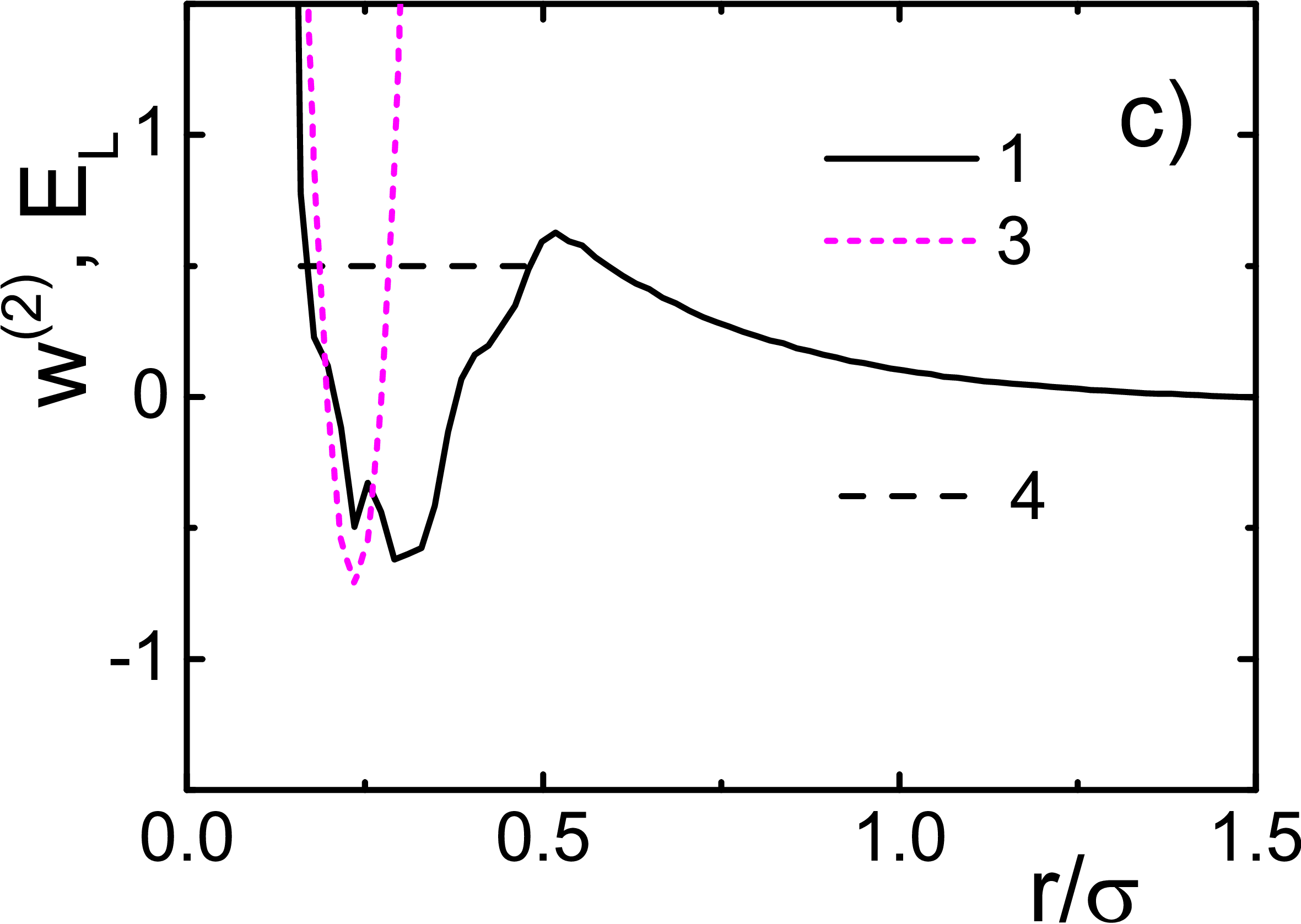}
	\includegraphics[width=0.8\columnwidth,clip=true]{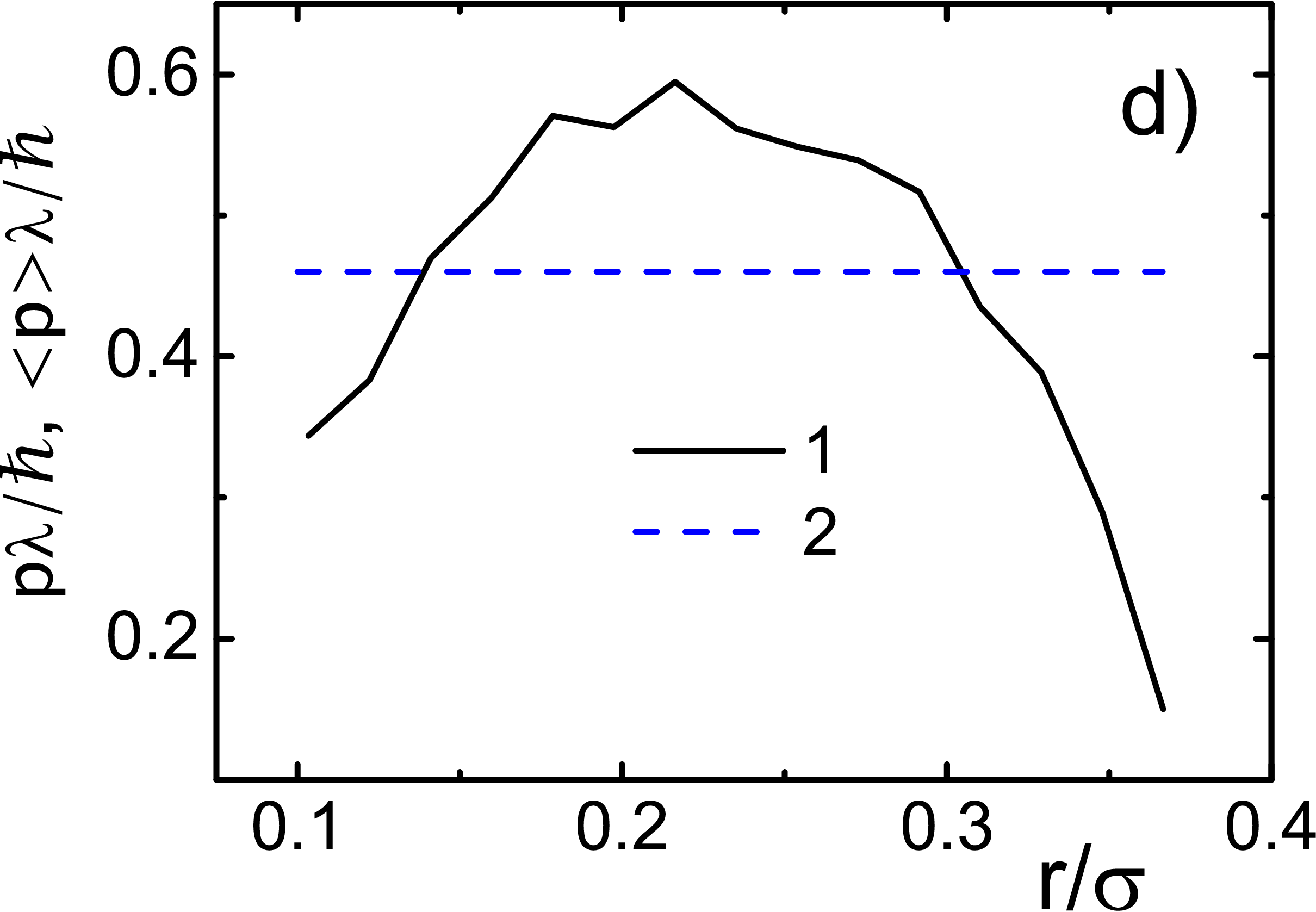}
	%\hspace{.2cm}
	\caption{(Color online)
		PMFs  $w^{(2)}(r/\sigma)$ for the same spin projections of fermions.
        Panels: a)---$n=0.6$; b)---$n=1.0$; c)---$n=1.4$;).
		Lines: 1---$L=0$; 2---$L=1$;
		3---pseudopotential well created by two neighboring fermions in the free space,
		4---energy level $E_L$ for $n_r=0$ and $L=0$; 5---$E_L$ for $n_r=0$ and $L=1$.
				Panel d). Typical fermion momenta in the Bohr--Sommerfeld condition.
		Lines: 1---$p(r)$ in (\ref{BS}); 2---$\langle p(r)\rangle$, average momentum for a bound state corresponding to $n=0.6$, $n_r=0$ and $L=0$.
		\label{dup}  
	}
\end{figure}

Let us consider the results of simulations. Lines 1 and 2 in panels a), b) and c) of Fig.~\ref{dup}
show the PMF $w^{(2)}(r)$ and the sum of $w^{(2)}(r)$ and $\frac{L(L+1/2)}{2r^2}$ for $L=1$ .
Line 3 in Fig.~\ref{dup} presents the pseudopotential well created by two neighboring 
%the same soft--spheres 
fermions in free space. Differences between $w^{(2)}(r)$ (line 1)  and the pseudopotential (line 3)  show the influence
of the medium on the pseudopotential well.

Let us use the Bohr--Sommerfeld condition \cite{nikiforov2005quantum} to determine possible
bound states in the spherically symmetric field $w^{(2)}$.
Our calculations according to (\ref{BS}) show the bound states for $n_r=0$ marked
by line 4 ($L=0$) in panels a), b), c) and 5 ($L=1$) in panel a) of Fig.~\ref{dup} for hardnesses $n=0.6, 1.0, 1.4$
respectively.  At $n\lsim 0.6 $ and  $n\gsim 1.4 $ the triplet bound states disappeared.

The momenta corresponding to these bound states can be taken as the average values of $<p(r)>$ in (\ref{BS})
(see lines 1 and 2 in panel d) of Fig.~\ref{dup}). The corresponding bound state momenta are also presented by circles 4 ($L=0$) and 5 ($L=1$)
in Fig.~\ref{pwp}. Let us stress that the positions of the peaks of the WPIMC MDF agree well enough with the positions of the circles
corresponding to the Bohr--Sommerfeld condition, supporting the correctness of our estimations.
%Fig.~\ref{pwp} demonstrates also the noticeable repetition in the position of the sharp similar peaks on the MDF
%for neighboring hardnesses at the same density and temperature.

%\begin{figure}[t] 

\section{Discussion}

Thermodynamic properties and RDFs
of the classical soft--sphere system have been studied many times in the literature
for the hardness of the potential larger than three (particularly, for $n = 12$, see the review  \cite{pieprzyk2014thermodynamic,branka2011pair}).
Here  we present the MDFs and RDFs obtained by WPIMC
for the hardness of the soft--sphere quantum pseudopotential smaller or of the order of unity ($n=0.2$, 0.6, 1.0, 1.4).
%by  the path integral Monte Carlo simulations in the Wigner formulation of quantum mechanics (WPIMC) .
%WPIMC calculations for larger hadness are in progress.

The obtained MDFs
demonstrate narrow sharp separated peaks disturbing the Maxwellian distribution.
The physical reason for this behavior is the following. The MDFs are
averaged over all fermions. The MDF of free fermions is Maxwellian
with the average momentum $p\lambda/\hbar \sim 2$ (see Fig.~\ref{pwp}).
As it follows from the Bohr--Sommerfeld condition the typical momentum of the bound state of a TFC
is $p\lambda /\hbar \sim 0.5$.
So the peaks on an MDF corresponding to the TFC bound states averaged over all particles
with the Maxwellian distribution %as the bound state peaks in the distribution 
will be smoothed by the thermal motion.
On the contrary, the momentum of slow free fermions with $p\lambda/\hbar \sim 1 $  is modified by
the momentum of the TFC bound states (of the order of $p\lambda/\hbar \sim 0.5  $) so sharp peaks appear
in the Maxwellian distribution as it is demonstrated in Figure~\ref{pwp}.

To summarize, let us note that the PIMC simulations in the Wigner approach to quantum mechanics used in this paper
allow us to calculate the MDFs and the spin--resolved RDFs
of a strongly correlated system of soft--sphere fermions for different hardness of the interparticle pseudopotential.
The obtained spin--resolved RDFs
%for different hardness of the potential 
demonstrate the appearance of three--fermion clusters (triplets)
caused by the interference of the exchange and interparticle interactions.
The semiclassical analysis in the framework of the Bohr--Sommerfeld quantization condition applied to the PMF corresponding to the same--spin RDF allows to detect the triplet exchange--correlation bound states
and to estimate the corresponding averaged energy levels.
%The noticeable repetition in the position of the series of similar sharp peaks on the MDFs for neighboring hardness at fixed density and temperature in 
%The independent WPIMC calculations confirms the reliability of the obtained results and average values of the energy levels. 

%\funding{
%	This research was supported by the 	Ministry of Science and Higher Education of the Russian Federation
%	(Agreement with Joint Institute for High Temperatures RAS No.075-15-2020-785 dated September 23, 2020).
%%(Fortov, судсидии)
%
%The theoretical approach to the basic equations and algorithmic realization  was supported 
%by the Russian Science Foundation, Grant No.~20-42-04421. 
%%%(RNF)
%
%The algorithmic realization of this approach was supported by the Ministry of Science and Higher Education of the Russian Federation 
%(Agreement with Joint Institute for High Temperatures RAS No 075-15-2020-785 dated September 23, 2020).  
%%(Fortov, судсидии)
%
%The numerical calculations were supported by the Ministry of Science and Higher Education
%of the Russian Federation (State Assignment No. 075-00460-21-00)
%NIR
%
%}
%
\section*{Acknowledgements}
We thank G.\,S.~Demyanov for comments and help in numerical matters. 
We value stimulating discussions with Prof. M. Bonitz, T. Schoof, S. Groth and T. Dornheim (Kiel).
% and J.W. Dufty and V. Karasiev (University of Florida). 
This research was supported by the Ministry of Science and Higher Education of the Russian Federation
(Agreement with Joint Institute for High Temperatures RAS No 075-15-2020-785 dated September 23, 2020).
%This work was supported by the Russian Science Foundation, Grant No. 20-42-04421. 
%This work has been supported by the Russian Science Foundation via grant 14-50-00124 and Deutsche Forschungsgemeinschaft 
%via SFB TR-24. Computations were performed at the ``Fermion'' compute-cluster of the Institute for Theoretical Physics 
%and Astrophysics of Kiel University and at the North-German Supercomputing Center (HLRN) via grant SHP006.
The authors acknowledge the JIHT RAS Supercomputer
Centre, the Joint Supercomputer Centre of the Russian
Academy of Sciences, and the Shared Resource Centre Far
Eastern Computing Resource IACP FEB RAS for providing
computing time.

%\section*{References}
%\bibliographystyle{iopart-num} 
%\bibliographystyle {tfo}
\bibliographystyle {apsrev}
%\bibliography{FilinLar}
%\bibliography{ueg.bib}
%%%%%%%\bibliography{ueg2020.bib} 

\end{document}